\documentstyle[psfig]{l-aa}

\newcommand{\kms}{km~s$^{-1}~$}

\begin{document}
\thesaurus{3
	(11.06.2;
	11.12.2;
	13.18.1)}

\title{On the local radio luminosity function of galaxies. II: environmental
dependences among late-type galaxies}
\author{G. Gavazzi\inst{1}\and
A. Boselli\inst{2}}
\offprints{Gavazzi
	   Gavazzi @ trane.uni.mi.astro.it}
\institute{Universit\'a degli Studi di Milano, Via Celoria 16, 20133, Milano, Italy 
\and
Laboratoire d'Astronomie Spatiale, Traverse du Siphon, BP 8, F-13376 Marseille Cedex 12, France}

\date{Received , accepted }
\maketitle
\markboth{the RLF of 5 galaxy clusters}{...}

\begin{abstract}

Using new extensive radio continuum surveys at 1.4 GHz (FIRST and NVSS), 
we derive the distribution of the radio/optical and radio/NIR luminosity (RLF) of late-type (Sa-Irr) galaxies
($m_p<15.7$) in 5 nearby clusters of galaxies: A262, Cancer, A1367, Coma and Virgo. 
With the aim of discussing possible environmental dependences of the
radio properties, we compare these results with those obtained for relatively isolated
objects in the Coma supercluster.
We find that the RLF of Cancer, A262 and Virgo are consistent with that of isolated
galaxies. Conversely we confirm earlier claims that galaxies in A1367 and Coma have their radio emissivity enhanced by a factor
$\sim 5$ with respect to isolated objects.
We discuss this result in the framework of the dynamical pressure suffered by
galaxies in motion through the intra-cluster gas (ram-pressure).
We find that the radio excess is statistically larger for galaxies in fast transit motion.
This is coherent with the idea that enhanced radio continuum activity is associated with
magnetic field compression. The X-ray luminosities and temperatures of Coma and A1367
imply that these two clusters have significantly larger intracluster gas
density than the remaining three studied ones, providing a clue for explaining the higher radio continuum luminosities of their galaxies. 
Multiple systems in the Coma supercluster bridge (with projected separations smaller 
than 300 kpc) have radio luminosities significantly larger than isolated galaxies.
\footnote{Table 1 is only available in electronic form at
the CDS via anonymous ftp to cdsarc.u-strasbg.fr (130.79.128.5)
or via http://cdsweb.u-strasbg.fr/Abstract.html}


\end{abstract}

\section{Introduction}

This is the second paper of a series aimed at studying the radio properties
of galaxies in the local universe. Paper I (Gavazzi \& Boselli 1998)
is devoted to a study of the dependence of the radio luminosity function
(RLF) on the
Hubble type and on the optical luminosity, using a deep radio/optical survey
of the Virgo cluster.\\
In the present paper we wish to discuss another issue:
is the local RLF of late-type galaxies universal or is it influenced 
by the environmental properties of galaxies?\\
Jaffe \& Perola (1976) found that late-type galaxies in some clusters 
(e.g. Coma) have unexpectedly overluminous radio
counterparts than "field" galaxies. Gavazzi \& Jaffe (1986) confirmed 
these early claims by
comparing the RLF of late-type galaxies within and outside rich clusters.\\
To re-address this question we derive in this paper the RLFs of galaxies 
in five nearby clusters and in less dense regions of the universe at 
similar distances. For this purpose we take advantage from the unprecedented 
homogeneous sky coverage of two recent
all sky radio surveys carried out with the VLA (NVSS and FIRST)
(Condon et al. 1998; White et al. 1997a).\\
Moreover precise photometric measurements became available in 
the regions under study.
For example a Near-Infrared (NIR) survey was
recently completed (Gavazzi \&
Boselli 1996; Gavazzi et al. in preparation; Boselli et al. in preparation), 
which is crucial for 
meaningful determinations of the radio properties of galaxies.
In fact the radio emission from spiral and irregular galaxies is to 
first order
proportional to their optical luminosity, as shown by Hummel (1981),
or to their mass, which is well traced by their NIR luminosity 
(Gavazzi et al. 1996).
Hence the necessity of properly normalizing the radio to the optical or NIR 
luminosities.\\
In Section 2 and 3 we discuss the optical sample used, the radio identification
procedure and the method for deriving the RLF.  
Differences in the
RLFs of the individual clusters (Section 4) are discussed in the framework of their
X-ray properties in Section 5.

\section{The Sample}

\subsection{The Optical Data}

The present investigation is based on the nearby clusters of galaxies 
A262, Cancer, Coma, A1367 and Virgo, and on relatively isolated 
objects in the Coma supercluster.\\
The optical sample is taken from the
CGCG Catalogue (Zwicky et al. 1961-68) in the regions:\\
$01^h43^m<\alpha<02^h01^m; 34.5^{\circ}<\delta<38.5^{\circ}$ (A262);\\ 
$08^h11^m<\alpha<08^h25^m; 20.5^{\circ}<\delta<23^{\circ}$ (Cancer) and\\ 
$11^h30^m<\alpha<13^h30^m; 18^{\circ}<\delta<32^{\circ}$ (Coma--A1367).\\
The latter region, beside the two rich clusters, contains
about 50\% of galaxies in relatively low density environments,
belonging to the bridge between Coma and A1367 (see Gavazzi et al. 1998). 
Within the limiting magnitude of the CGCG ($m_p\leq $15.7) these regions
contain 448 late-type (Sa-Irr) galaxies, all (but 3) with a redshift 
measurement in the literature (see Gavazzi et al. 1998).
Coordinates are measured with 1-2 arcsec error.
Photographic photometry
with an uncertainty of $\sim$ 0.35 mag is given in the CGCG.
Near Infrared (H) photometry is also available for all galaxies
except 5 (see Gavazzi \& Boselli 1996 and Gavazzi et al. in preparation).\\
The Virgo sample, extracted from the VCC (Binggeli et al. 1985)
is fully described in Paper I. Here we use a subsample of 174 late-type objects 
limited to $m_B\leq $14.0. The H band magnitudes are from  
Gavazzi \& Boselli (1996) and Gavazzi et al. (in preparation).

\subsection{1.4 GHz continuum data}

Radio continuum 1.4 GHz data in the regions covered by the present 
investigation are available from a variety of sources:\\
1) Full synthesys and snap-shot observations of specific regions 
were undertaken with the VLA and with the WSRT ("pointed" observations).
Jaffe \& Gavazzi (1986), del Castillo et al. (1988) and 
Gavazzi \& Contursi (1994) observed with the VLA several regions of the 
Coma supercluster. Venturi et al. (1990) took similar data of the Coma 
cluster with the WSRT. Bravo Alfaro (1997) derived some continuum measurements 
of galaxies in the Coma cluster from his VLA 21 cm line survey.
Gioia \& Fabbiano (1986), Condon (1987) and Condon et al. (1990) 
observed with the VLA relatively nearby galaxies projected onto the 
Coma regions. Salpeter \& Dickey (1987) carried out a survey of the 
Cancer cluster with the VLA. These surveys do not generally constitue a 
complete set of observations.\\
2) Recently, two all-sky surveys carried out with the VLA at 1.4 GHz became 
available:\\
a) the B array (FWHM = 5.4 arcsec) FIRST survey (1997 release) covers the sky north of 
$\delta >22^{\circ}$, with an average rms=0.15 mJy (White et al. 1997a).\\
b) the D array (FWHM = 45 arcsec) NVSS survey covers the sky north of 
$\delta >-40^{\circ}$,  with an average rms=0.45 mJy (Condon et al. 1998).
Except in specific regions
of the sky near bright sources, where the local rms is higher than average, 
these surveys
offer an unprecedented homogeneous sky coverage. They not only provide us 
with extensive 
catalogues of faint radio sources, but also with homogeneous upper limits 
at any celestial position.\\
Since radio data from more than one source exist for several target galaxies,
we choose between them adopting the following list of priority:\\
1) in general we prefer NVSS data to any other source because of its 
homogeneous character,
relatively low flux density limit and because its FWHM beam better 
matches the 
apparent sizes of galaxies under study, thus providing us with flux 
estimates little affected by missing extended flux.\\
2) For individual bright radio galaxies (eg. M87, N3862, N4874) we prefer
data from specific "pointed" observations since they should provide us 
with more reliable estimates of their total flux.\\
3) in all cases where the flux densities from NVSS are lower than those 
given in other references
we privilege the reference carrying the highest flux density.\\
4) in the region of the Coma superluster north of 22$^{\circ}$ we made
a comparison between the flux measurements derived from all available 
surveys (including FIRST).
As expected, the NVSS flux densities are systematically 1.9 times larger 
than the FIRST ones. 
Furthermore several NVSS sources are undetected in the FIRST data-base. 
These correspond
to slightly extended sources resolved by the FIRST beam, thus with peak 
flux density lower than the survey limit.
Conversely it seldom happens that FIRST sources are undetected in the 
NVSS data-base.
These are faint compact sources below the NVSS limiting flux density. 
In both cases the detections are often confirmed by
independent "pointed" measurements. Thus we assume the reference carrying the 
largest flux density. 

\subsection{The radio-optical identifications}

At the position of all optically selected galaxies 
we search for
a radio-optical coincidence. For the remaining undetected galaxies 
we compute an upper limit flux using $4\times rms$.
For this purpose we proceed as in Paper I with the following modifications:\\
\noindent
1) In the Coma supercluster region ($\delta >22^{\circ}$) 
we pre-select sources from the FIRST data-base with a maximum radio-optical 
positional discrepancy of 30 arcsec from the target galaxies.\\
\noindent
2) In all regions  
we pre-select sources from the NVSS data-base, allowing for a maximum 
radio-optical positional discrepancy of 30 arcsec.\\
3) at the position of all pre-selected optical-radio matches we compute an 
"identification class" (ID) according to Paper I.
\noindent

\begin{figure*}
\vbox{\null\vskip 16.0cm
\includegraphics{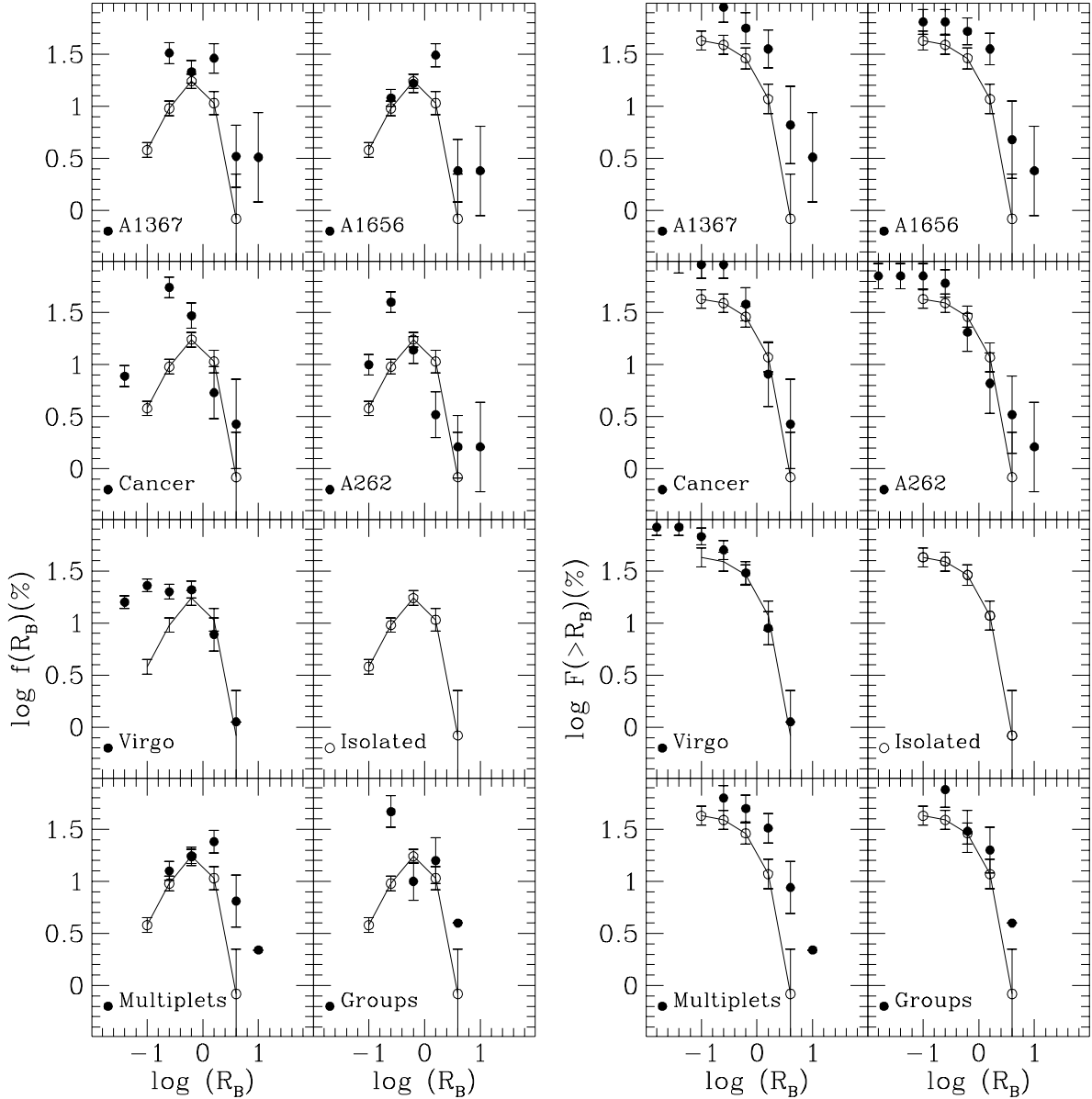}
}
\caption{the differential (a) and cumulative (b) RLFs as a function of the radio/optical ratio $R_B$ for late-type galaxies in 5 clusters and for isolated, multiplets and groups
in the Coma supercluster. The RLF of isolated (open dotes) is given in all other panels
for comparison.
} 
\label{Fig.1}
\end{figure*}

\begin{figure*}
\vbox{\null\vskip 16.0cm
\includegraphics{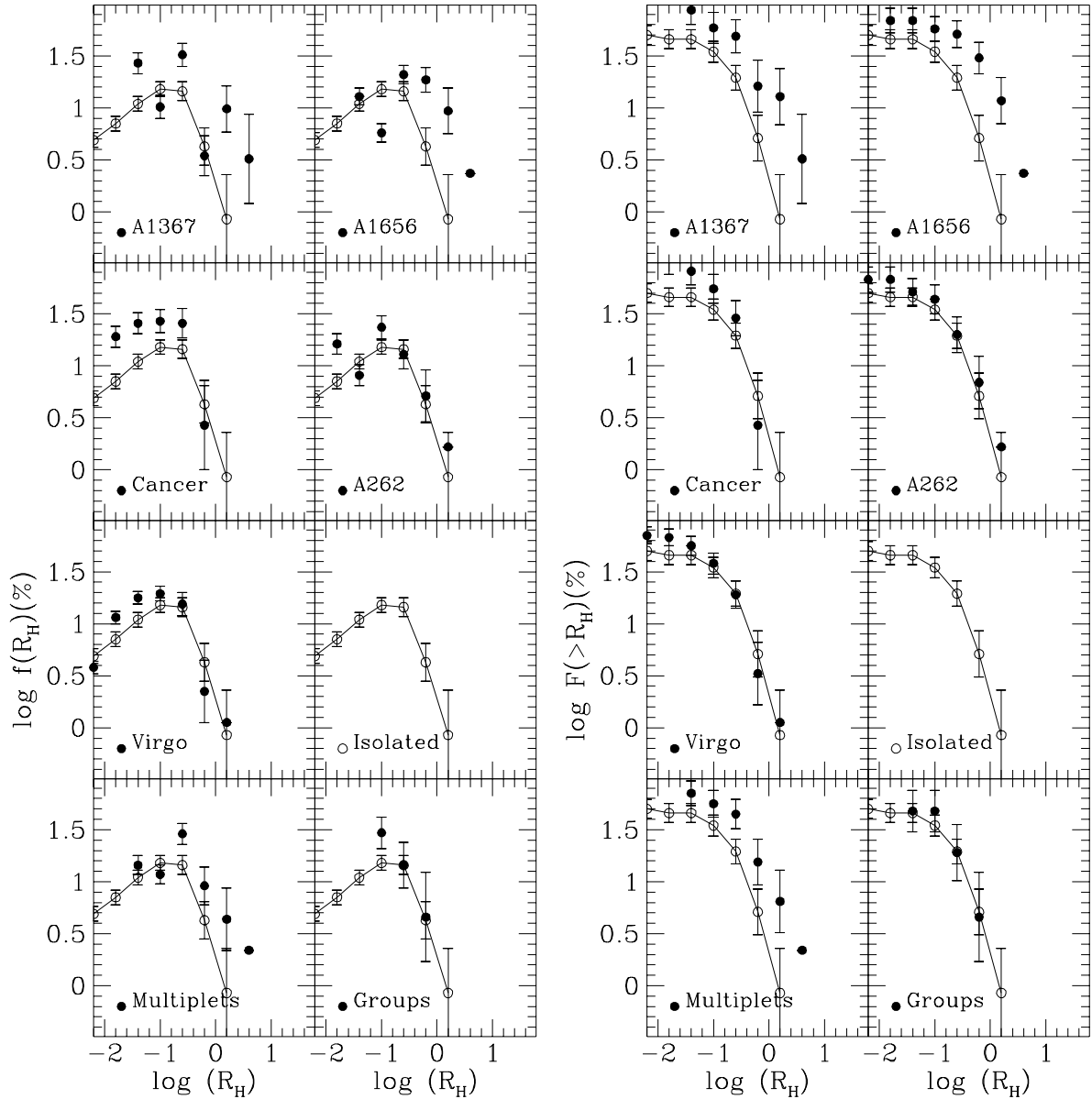}
}
\caption{the differential (a) and cumulative (b) RLFs as a function of the radio/NIR ratio $R_H$ for late-type galaxies in 5 clusters and for isolated, multiplets and groups
in the Coma supercluster. The RLF of isolated (open dotes) is given in all other panels
for comparison.
} 
\label{Fig.2}
\end{figure*}

The 408 positive radio-optical matches are listed in Table 1 as follows: \newline

Column 1: the CGCG (Zwicky et al. 1961-68) designation.\newline
Column 2: the photographic magnitude corrected for extinction in
our Galaxy according to Burstein \& Heiles (1982) and for internal extinction
following the prescriptions of Gavazzi \& Boselli (1996). \\
Column 3: the H band (1.65 $\mu m$) magnitude corrected for internal 
extinction following the prescriptions of Gavazzi \& Boselli (1996). \\
Column 4: the morphological classification.\\
Column 5: the membership to the individual clusters and clouds as 
defined in Gavazzi et al. (1998).\\
Columns 6, 7: the (B1950) optical celestial coordinates of the target galaxy.\\
Columns 8, 9: the (B1950) celestial coordinates of the radio source.\\
Column 10: the radio-optical offset (arcsec).\\
Columns 11: the identification class (see Paper I for details).
ID=1 and 2 are good identifications. ID=4 correspond to radio sources projected
within the galaxy optical extent. ID=3 are dubious identifications not used in
the following analysis.\\
Column 12: the 1.4 GHz total flux density (mJy).\\
Columns 13, 14: the extension parameters of the radio source (major and minor axes
in arcsec).\\
Column 15: reference to the 1.4 GHz data.\\
All sources listed in Table 1 are found within 30 arcsec from the central 
optical coordinates of the parent galaxies.  
An estimate of the number of possible chance-identifications ($N_{c.i.}$) 
among the 408 sources/galaxies listed in Table 1 is carried out using Condon 
et al. (1998) Fig. 6. The probability 
of finding an unrelated source within 30 arcsec of an arbitrary position
is 1~\%. Thus about 4 sources in Table 1 should be spurious associations.

\section{The RLF}

Spiral galaxies are well known to develop radio sources with an average 
radio luminosity proportional to their optical luminosity (see Paper I). 
For these objects it is convenient
to define the (distance independent) radio/optical ratio: 
$R_\lambda= S_{1.4} / k(\lambda) 10^{-0.4*m(\lambda)}$,
where $m(\lambda)$ is the magnitude at some wavelength $\lambda$.
$k(\lambda)=4.44\times 10^6$ and $1.03 \times 10^6$ are the factors 
appropriate
to transform in mJy the broad-band B and H magnitudes respectively.
$R_B$ gives the ratio of the radio emission per unit light emitted 
by the relatively
young stellar population, while the Near Infrared $R_H$ gives the ratio 
of the radio emission per unit 
light emitted by the old stellar population, thus per unit dynamical 
mass of the system (see Gavazzi 1993; Gavazzi et al. 1996).
The Fractional Radio Luminosity Function (RLF), that is the 
probability distribution 
$f(R_\lambda)$ that galaxies develop a radio source of a given radio/optical 
ratio $R_\lambda$,
from a complete, optically selected sample of galaxies is derived  
using equation (5) of Paper I.

\section{Results}

\subsection{The environmental dependence of the RLF}

The analysis in this section is aimed at determining whether the radio 
properties of 
late-type (S-Irr) galaxies depend on their environmental conditions. 
For this purpose we compare the RLFs of galaxies in 5 rich clusters with 
those of galaxies belonging to the relatively isolated regions of the Coma 
supercluster. 
According to the definition of Gavazzi et al. (1998), who studied the 
3-D distribution of galaxies in this supercluster, galaxies with no 
companion within 0.3 Mpc projected radius can be considered "isolated"; 
"multiplets" 
have at least one companion within 0.3 Mpc projected radius
and within 600 \kms, and "groups" have at least 8 galaxies within 0.9 
Mpc projected radius and within 600 \kms.\\
We derive for these objects the differential frequency distributions 
$f(R_H)$ and $f(R_B)$ and the
corresponding integral distributions $F(\geq R_H)$ and $F(\geq R_B)$, 
binned in intervals
of $\Delta R_\lambda = 0.4$. These are shown in Fig. 1 and 2, a and b 
respectively.\\
a) The shape of the differential $f(R_B)$s is typical of a
normal distribution peaked at log $(R_B)$ between -0.5 and 0. This confirms 
that there is a
direct proportionality between the mean radio and optical luminosities. 
About 15 \% of all galaxies are detected at the peak of the distribution. 
About 50\% of all galaxies
have a radio/optical ratio greater than 0.01.\\
b) The shape of the differential $f(R_H)$s is similar to the $f(R_B)$s, 
but with a somewhat
larger dispersion. This confirms that the radio luminosity better 
correlates with the young than with the old stellar population.\\
c) It appears that both $f(R_H)$ and $f(R_B)$ of isolated galaxies, 
mambers of groups and of the Cancer, 
A262 and Virgo clusters are statistically consistent.
Moreover we searched for possible differences among the various 
subclusters within the Virgo cluster (cluster A, B, southern extension) 
and found none. These results confirm that 
the RLFs of the Virgo, Cancer and A262 clusters are similar to the 
field one, as claimed by
Kotanyi (1980), Perola et al. (1980) and by Fanti et al. (1982), 
resepectively.\\
d) The RLFs of the Coma and A1367 clusters and the Coma supercluster 
multiplets, on the contrary, show significantly enhanced radio emission: 
at any given log $R_\lambda$ above -0.5, the probability of finding a 
radio source
associated with a late-type galaxy in these clusters is a factor of 
$\sim$ 10 higher than for other
galaxies. Conversely at fixed $f(R_\lambda)$ (i.e. 10 \%) these galaxies 
have the
ratio $R_\lambda$ a factor of 5 higher than for the remaining galaxies.
The overluminosity of Coma with respect to the "field" was claimed by 
Jaffe \& Perola (1976), later confirmed by Gavazzi \& Jaffe (1986). Similar 
evidence was found for A1367 by Gavazzi (1979) and confirmed by 
Gavazzi \& Contursi (1994).\\
Evidences c) and d) are even more clear-cut in the cumulative distributions. 
However, the reader should remember that cumulative distributions tend to 
emphasize differences if they are present in the highest $R_\lambda$ bin.
Fig. 3 shows that the cumulative fraction of galaxies with $F(> R_B=0.2)$
is consistently below 10\% for isolated galaxies, members of A262, 
Cancer and Virgo, and consistently above 30\% for multiplets and members of A1367 
and Coma.

\begin{figure*}
\vbox{\null\vskip 7.5cm
\includegraphics{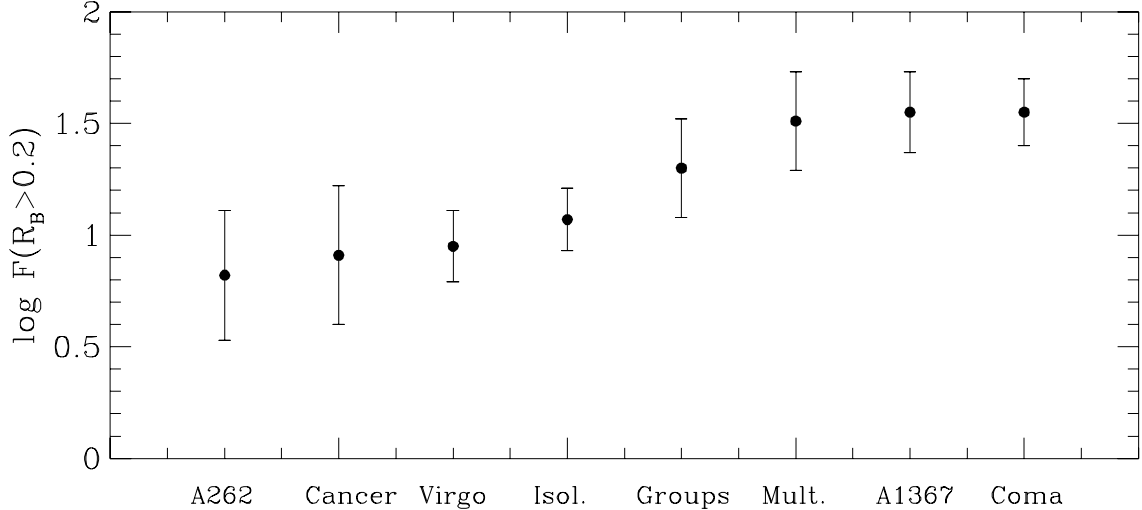}
}
\caption{ the cumulative RLF for $log R_B >~0.2$ is given for the various clusters and
substructures.
} 
\label{Fig.3}
\end{figure*}

\section{Discussion and Conclusions}

\subsection{Cluster Galaxies}

We have shown (Section 4.1) that late-type galaxies in the
clusters A1367 and Coma develop radio sources more frequently
than galaxies 
in the remaining clusters or more isolated galaxies.
Here we wish to discuss if this evidence is connected with
the properties of the hot gas permeating the clusters (IGM), 
which might emphasize the role of the environment.\\  
Enhanced radio emission and morphological disturbances 
both in the radio and in the optical have been observed in three
Irr galaxies in A1367 (CGCG 97073, 97079 and 98087) showing radio trails exceeding
50 kpc in length (see Gavazzi \& Jaffe 1985; 
Gavazzi et al. 1995). A highly asymmetrical HI structure has been reported in
NGC 4654 in the Virgo cluster (Phookun \& Mundy, 1995) and several other
examples are discussed in Gavazzi (1989).
These peculiarities have been interpreted
in the ram-pressure scenario: galaxies in fast motion 
through the inter-galactic medium experience enough dynamical pressure
to compress their magnetosphere on the up-stream side, form
a tail-like radio structure on the down-stream side and produce 
a net enhancement of the radio luminosity. 
These galaxies should have experienced such a pressure for a relatively
short time, otherwise their HI content would have been strongly reduced
by stripping, contrary to the observations.
A similar interpretation has been proposed to explain the asymmetries in
NGC 1961 (Lisenfeld et al. 1998) and NGC 2276 (Hummel \& Beck 1995).
In these cases however the gravitational interaction with companions provides an 
alternative interpretation of the observed asymmetry (see Davis et al. 1997).  
Although 
these phenomena have not been observed in the Coma cluster, perhaps due
to the lack of appropriate sensitivity/resolution, it cannot be excluded 
that galaxies in this cluster have radio luminosities enhanced
by the same mechanism.\\
It is in fact remarkable that the radio/H ratio $R_H$ of the detected 
galaxies indicates that the radio emissivity increases with
the transit velocity through the IGM.\\
Fig. 4 shows $log R_H$ as a function of the deviation of the 
individual velocities (projected along the line of sight) 
from the mean velocity of the cluster to which they belong.
Highly HI deficient objects ($Def_{HI}>0.8$) are excluded from the plot because
for these galaxies the star
formation rate, thus the radio emissivity, might have been totally quenched by the dynamical pressure, due to complete gas removal (see Cayatte et al. 1990).
Galaxies populate a "wedge" region in the $R_H$ vs. 
$|\Delta V|=|V_g - <V_{cl}>|$ plane.
This is because some galaxies in fast transverse motion through the cluster 
might appear at low $|\Delta V|$ if their motion is parallel to the plane 
of the sky. For example the 50 kpc long radio trails associated with
three A1367 galaxies mentioned above testifies that a significant component of
their velocity lies in a plane perpendicular to the line of sight.
The wedge pattern is observed in all clusters, but to a lesser degree in Virgo.
Fig. 4. also reports $<log R_H>$ averaged below and above $|\Delta V|=720~km~s^{-1}$,
showing that the average contrast between the radio emissivity at low vs. high $|\Delta V|$
ranges between a factor of 2 and 7, with a mean of 3. This evidence by itself is sufficient
to rule out that enhanced radio activity is associated with galaxy-galaxy 
interactions, which is expected to be more effective at small velocity differences.\\
An estimate of the average dynamical pressure:
$P_{ram} \sim n_e \times \Delta V^2$
can be derived from the global cluster X-ray luminosity and temperature:
$n_e^2 = k L_x/T_x^{1/2}$.
Adopting $L_x$ and $T_x$ for Virgo, A262, A1367 and Coma from the recent compilation by White et al. (1997b) (their Table 1) and for Cancer from Trinchieri (1997)
and adopting the velocity dispersions of the
individual clusters (taken from White et al. 1997b and from our own data
for Cancer) we compute that the effective dynamical pressure experienced by
galaxies in the individual clusters is nearly absent in the Cancer cluster, 10 times 
higher in A262 and Virgo, 30 times higher in A1367 and 300 times higher in the Coma cluster,
providing a hint to explain the excess radio emission in A1367 and Coma with respect
to all other clusters.
 
\begin{figure*}
\vbox{\null\vskip 16.0cm
\includegraphics{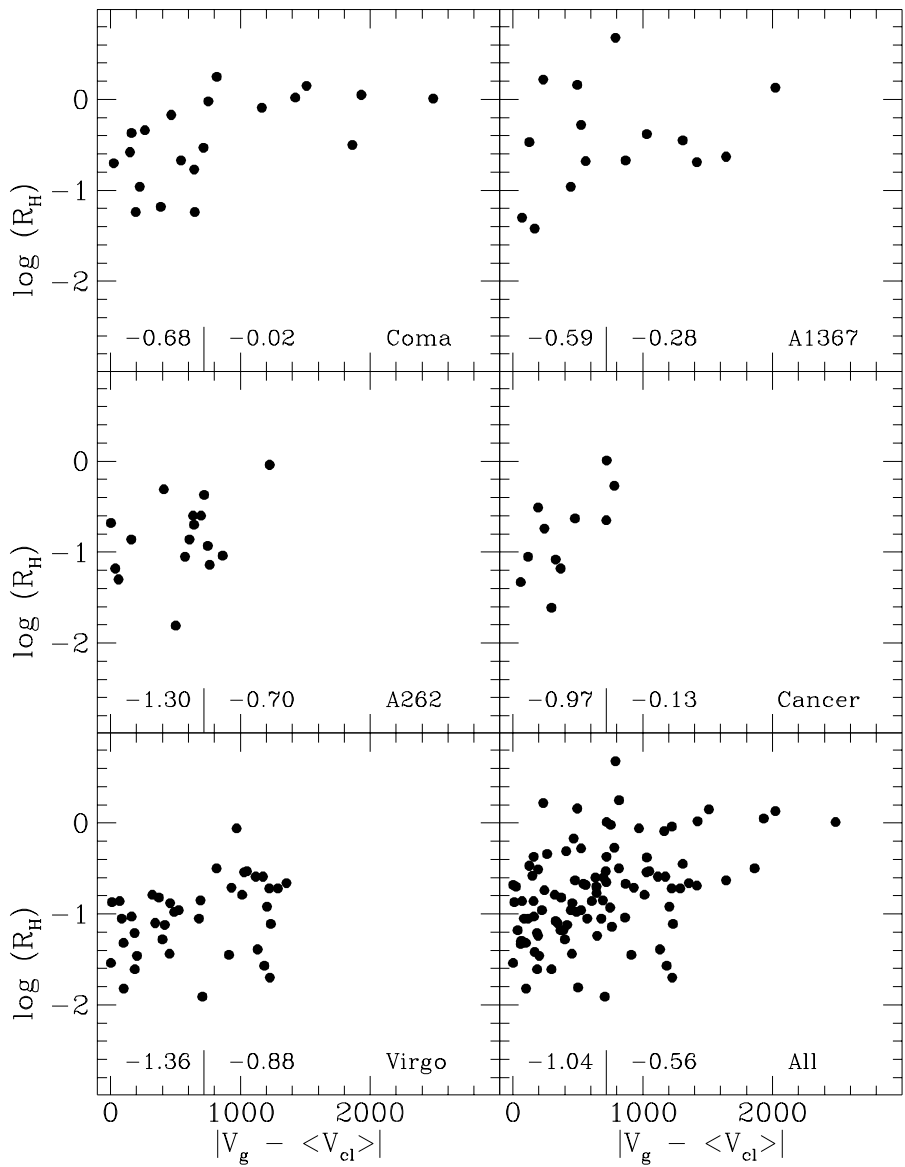}
}
\caption{the distribution of the NIR $R_H$ as a function of the deviation (along the
line of sight) of the individual velocities from the cluster average velocity. The plot
includes the detected galaxies separately for the 5 clusters and grouped all together
(bottom-right panel). The values of $<log R_H>$ averaged below and above $|\Delta V|=720~km~ s^{-1}$ are given in each panel.} 
\label{Fig.4}
\end{figure*}

\subsection{Multiplets}

Also multiple systems in the Coma supercluster bridge have radio/optical
ratios significantly larger than isolated galaxies, suggestive of an enhanced
star formation rate in galaxies showing some degree of interaction (our multiplets
have projected separations lower than 300 kpc).
Hummel et al. (1990) compared the central (within 10 arcsec from the nuclei) radio luminosity
of isolated galaxies with those of double systems with average separations 4-5
effective radii. They found that
the central radio sources in interacting spiral galaxies are on average 
a factor of 5 stronger than the ones in the more isolated galaxies.
This difference is almost completely due to the activity in the HII 
region nuclei.
Menon (1995) analyzed the radio luminosity of spiral/Irr galaxies in Hickson Compact Groups (HCG).
He found that the radio 
radiation from the nuclear regions is more than 10 times that from comparable 
regions in the comparison sample, but that the extended radiation from HCG spirals is 
lower than from a comparison sample of isolated galaxies.
This evidence is interpreted in 
a scenario in which galaxy interactions cause massive inflows 
of gas towards the centers. The resulting
star formation in the center leads to the formation of supernovae and
subsequent radio radiation.
Our radio observations have a mix of resolutions (from about 5 to 45 arcsec)
which unfortunately do not allow us to resolve homogeneously
the contribution of nuclear sources to the total flux.
We must conclude that, to the first order, the radio emissivity
in our sample of multiplets is dominated by the enhanced nuclear activity.

\section{Summary}

In summary the present investigation brought us to the following empirical results:\\
1) the RLF of Cancer, A262 and Virgo are consistent with that of isolated galaxies. \\
2) galaxies in A1367 and Coma have their radio emissivity enhanced by a factor
$\sim 5$ with respect to isolated objects.
We find that the radio excess is statistically larger for cluster galaxies with
large velocity deviations with respect to the average cluster velocity. 
This is coherent with the idea that enhanced radio continuum activity is produced by
magnetic field compression of galaxies in fast transit motion through the intra-cluster gas,
and is inconsistent with the hypothesis that the phenomenon is due to galaxy-galaxy interactions.
The higher X-ray luminosies and temperatures of Coma and A1367 compared with the 
remaining three studied clusters provides a clue for explaining why the radio enhancement
is observed primarely in these two clusters. \\
3) Multiple systems in the Coma supercluster bridge (with projected separations smaller 
than 300 kpc) have radio/optical
ratios significantly larger than isolated galaxies, suggestive of an enhanced
star formation rate probably taking place in the nuclear regions.

\acknowledgements {We wish to thank P. Pedotti for her contribution 
to this work and T. Maccacaro for useful discussions.
We wish also to acknowledge the NVSS and FIRST teams for their
magnificent work.}

\end{document}